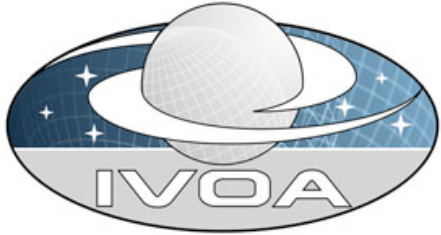

# Simple Image Access Specification Version 1.0

## IVOA Recommendation 2009-11-16

**Author(s):**
  Doug Tody (NRAO)
  Ray Plante (NCSA)
**Editor:**
  Paul Harrison (AstroGrid)

## Abstract

This specification defines a protocol for retrieving image data from a variety of astronomical image repositories through a uniform interface. The interface is meant to be reasonably simple to implement by service providers. A query defining a rectangular region on the sky is used to query for candidate images. The service returns a list of candidate images formatted as a VOTable. For each candidate image an access reference URL may be used to retrieve the image. Images may be returned in a variety of formats including FITS and various graphics formats. Referenced images are often computed on the fly, e.g., as cutouts from larger images.

## Status of this Document

This document has been reviewed by IVOA Members and other interested parties, and has been endorsed by the IVOA Executive Committee as an IVOA Recommendation. It is a stable document and may be used as reference material or cited as a normative reference from another document. IVOA's role in making the Recommendation is to draw attention to the specification and to promote its widespread deployment. This enhances the functionality and interoperability inside the Astronomical Community.

The first release of this document was 2002 September 30. Minor changes were made for the IVOA version prepared 2004 May 15. More substantial changes were made in October 2008 to prepare for Recommendation - these changes were made to reflect the fact that the Simple Image Access Protocol had been a *de facto* standard for some time, but that parts of the document did not any longer reflect the established use.

A list of current IVOA Recommendations and other technical documents can be found at http://www.ivoa.net/Documents/.

# Contents

---

## 1. Overview

This specification defines a prototype standard for retrieving image data from a variety of astronomical image repositories through a uniform interface. The interface is meant to be reasonably simple to implement by curators of image repositories.

This specification is based primarily on two documents. The first document, "Simple Image Retrieval: Interface Concepts and Issues", describes a longer term view of how simple image access can fit into a more general framework for image access in the VO. The URL-based implementation specified here is intended to be consistent with the concepts discussed in this document. The second document, the "Simple Cone Search specification" provides a means to query catalogs via HTTP with a uniform interface. A large number of implementations of the Cone Search specification have demonstrated that a simple standard based on name-value HTTP GET requests and the VOTable XML format is both easy to implement and useful for catalog data access. The Simple Image Access interface (SIA) defined here follows a similar approach.

SIA is designed primarily as an "image on demand" service, with images created on-the-fly by the service given the position and size of the desired output image as specified by the client. In the ideal case the image service presents an image data collection to a client as a seamless *virtual sky*, allowing the client to "observe" any region within the coverage of the service, without having to worry about details such as image boundaries or calibration. SIA can also be used to search for and retrieve pre-existing, statically sized atlas or pointed images which overlap the given search region on the sky. When "observing" the virtual sky, image query and access is normally limited to regions small enough to be represented as a single image. Larger regions can be queried when performing a single search of a static image data collection. Data collections are often distributed, and the client may query multiple image services simultaneously, e.g., to gather data from multiple wavelength regimes or surveys to analyze a single region on the sky.

In operation SIA represents a negotiation between the client and the image service. The client describes the ideal image - what it would like to get back from the image service - and the image service returns a list, encoded as a VOTable, of the (often virtual) images it can actually return. The client then examines this to determine if it is interested in any of the available images, possibly iterates with the service to refine the query, then issues a series of getImage requests to retrieve the selected images. A key point is that it is entirely up to the image service what images, if any, it offers to the client. These images may range from a simple list of static archive images which intersect the region of interest defined by the client, to a mosaiced and reprojected synthetic image matching the ideal image requested by the client. In the latter case the image is not computed until it is actually accessed or retrieved by the client.

The bulk of this document represents a technical specification for the simple image access interface. For examples of how the interface is intended to be used, please refer to the Usage Examples section below.

## 2. Requirements for Compliance

Compliance with this specification requires that an image access web service be maintained with the following characteristics:

1. The **Image Query** web method **must** be supported as defined in section 4 below.

   Through this web method, clients search for available images that match certain client-specified criteria. The response is a table that describes the available images, including image metadata and access references (implemented here as URLs) for retrieving them. Images thus referenced may be *real* (pre-existing static images) or *virtual* (images generated on-the-fly when accessed).

2. The **Image Retrieval** (getImage) web method **must** be supported as defined in section 5 below.

   This method allows clients to retrieve single images through a simple synchronous HTTP GET request using the access reference (URL) returned earlier by the image query. The response to the GET request is the requested image, returned with a MIME type such as "image/fits", "image/jpeg", and so forth. Both FITS and graphics images (JPEG, PNG, etc.) can be retrieved. Any FITS images returned by the service should contain a valid FITS world coordinate system (WCS).

3. Finally the image service **should** be **registered** by providing the information defined in section 6 below.

   Registration allows clients to use a central registry service to locate compliant image services and select an optimal subset of services to query, based on the characteristics of each service and the image data collections it serves.

Here, **image service** refers to a web service which returns images or image metadata to a client. The Simple Image Access specification defines a number of *web methods* which services must provide to implement a compliant image service. A **web method** is a function defined by a web service and called via the web. Since the Simple Image Access service described here is URL-based (e.g., as opposed to using SOAP/WSDL), the web methods described here are implemented as URLs using HTTP GET or POST request/response patterns as specified by the HTTP Standard.

### 2.1 Compliance

The keywords "**must**", "**required**", "**should**", and "**may**" as used in this document are to be interpreted as described in RFC 2119.

An implementation is compliant if it satisfies all the **must** or **required** level requirements for the protocols it implements. An implementation that satisfies all the **must** or **required** level and all the **should** level requirements for its protocols is said to be "unconditionally compliant"; one that satisfies all the **must** level requirements but not all the **should** level requirements for its protocols is said to be "conditionally compliant".

## 3. Image Service Types

It is assumed that compliant image services fall into one of four categories. These categories are used primarily to characterize the service as part of service registration, to aid clients in discovering the services best suited to their needs. How an image service implements this specification depends on which category it falls in. The categories are defined as follows:

1. **Image Cutout Service**
   This is a service which extracts or "cuts out" rectangular regions of some larger image, returning an image of the requested size to the client. Such images are usually drawn from a database or a collection of survey images that cover some large portion of the sky. To be considered a cutout service, the returned image should closely approximate (or at least not exceed) the size of the requested region; however, a cutout service will not normally resample (rescale or reproject) the pixel data. A cutout service may mosaic image segments to cover a large region but is still considered a cutout service if it does not resample the data. Image cutout services are fast and avoid image degradation due to resampling.

2. **Image Mosaicing Service**
   This service is similar to the image cutout service but adds the capability to compute an image of the size, scale, and projection specified by the client. Mosaic services include services which resample and reproject existing image data, as well as services which generate pixels from some more fundamental dataset, e.g., a high energy event list or a radio astronomy measurement set. Image mosaics can be expensive to generate for large regions but they make it easier for the client to overlay image data from different sources. Image mosaicing services which resample already pixelated data will degrade the data slightly, unlike the simpler cutout service which returns the data unchanged.

3. **Atlas Image Archive**
   This category of service provides access to pre-computed images that make up a survey of some large portion of the sky. The service, however, is not capable of dynamically cutting out requested regions, and the size of atlas images is predetermined by the survey. Atlas images may range in size from small cutouts of extended objects to large calibrated survey data frames.

4. **Pointed Image Archive**
   This category of service provides access to collections of images of many small, "pointed" regions of the sky. "Pointed" images normally focus on specific sources in the sky as opposed to being part of a sky survey. This type of service usually applies to instrumental archives from observatories with guest observer programs (e.g., the HST archive) and other general purpose image archives (e.g., the ADIL). If a service provides access to both survey and pointed images, then it should be considered a Pointed Image Archive for the purposes of this specification; if a differentiation between the types of data is desired the pointed and survey data collections should be registered as separate image services.

The **image data model** assumed here is minimal at this point. An image should be a calibrated object frame imaging some region of the sky. Only two dimensional images are fully supported within the interface at this time, but an actual external FITS file can be of higher dimension provided the two spatial axes are mapped into the view seen here (e.g., a spectral data cube should be presented as a 2D image within SIA even though the actual external FITS file may have 3 or more image dimensions). Within the current interface, images consist of a pixel array, an image header describing the image, and optionally an associated world coordinate system and spectral bandpass data model. The actual image file may contain additional information not defined here.

Images can be returned as either FITS files or as graphics images. For the FITS format, each image file thus returned should contain a single image in the primary header unit (PHU). Extensions are permitted, but their usage is not defined by this interface, and the service should assume that an external client will only access the PHU. Image aggregates are permissible but are not explicitly supported by the interface. For example, a calibrated mosaic detector observation stored as a FITS multi-extension file (MEF) should be presented to the outside world as a set of separate but related images, even if the images are stored in the archive as extensions within a single FITS MEF file. In such a case the service would need to extract each image from the MEF to return it to the client at access time. Service-defined keywords or table columns can be used to associate image aggregates of various types.

Ultimately, **VO data models** will provide a means to describe more complex data objects within the VO than be directly addressed by the SIA prototype. VO data models are needed for many data objects or attributes such as the limiting magnitude/flux/SNR of an image, for the spatial resolution of an image (characteristic PSF), for the temporal coverage of an image, for any data quality, exposure time, or variance masks associated with a pixel array, and so forth.

# 4. Image Query

The purpose of the **image query web method** is to allow users to search for image data for a given region on the sky. Additional parameters may optionally be used to further refine the query. The size, scale, projection, and so forth of the ideal output image can be specified, and the service will return references to images which most closely match what is requested. How closely the service can match the specifications of the image requested by the client will depend upon the type of image service as defined in section 3 above. The service returns a table listing all available images satisfying the query. Descriptive metadata is returned for each image to allow the client to decide which, if any, images it wishes to retrieve. An access reference is provided for each image to permit subsequent staging and retrieval.

The maximum size of the **region of interest** for a query depends upon the type of image service. Image cutout and mosaicing services are designed to generate a single image covering the region of interest specified by the client. Unless the desired image resolution is very low, the largest region on the sky that it is practical to cover with a single image is typically on the order of a degree or so (a 1 degree field at a resolution of 0.5 arcsec/pixel requires an image 7200 pixels square, or 104 MB at 16 bits per pixel). Larger images are certainly possible, but are not very practical for a service which is generating images on the fly and delivering them to a remote client over the Internet. A more typical image cutout is likely to be much smaller, on the order of several hundred pixels square or less, with analysis involving many such small regions. If a client wants to use a cutout or mosaicing service to cover a very large region at high resolution it would need to make multiple requests to generate and download all the desired image data.

Unlike the cutout and mosaicing services, image services which merely return lists of pre-existing atlas or pointed images can potentially support very large search regions in queries. The same interface is used in both cases, with the maximum (useful) size of the region of interest for the image query being specified in the service registration parameters. Our expectation is that most image queries will involve modest sized query regions.

Mosaicing services are sophisticated enough to be able to generate images on the fly given the field center, scale, rotation, projection, and so forth specified by the client. To support such services we need a way to allow the client to specify the parameters of the ideal image. This is done using a set of **image generation parameters**. These are optional - a client is always free to specify them, but if the image service doesn't support on-the-fly image generation it is free to ignore them and merely describe the images, if any, that it can provide. The default values for the optional image generation parameters are derived from the region of interest used for the image query. If the image generation parameters are fully specified these override the defaults provided by the region of interest.

## 4.1 Input

1. The image query input is sent via HTTP GET method to a URL which conforms to standard URI syntax as defined by the IETF RFC 3986. This URL may be considered to have two parts:
   1. A base URL, which defines the address of the service, of the form:
      **http://**_<server-address>_**/**_<path>_**?**_[<extra GET arg>_**&**_[...]]_

      > **Examples:**
      > ```
      > http://myimages.org/cgi-bin/VOimq?
      > http://adil.ncsa.uiuc.edu/voimagequery?resolve&issurvey=T&
      > ```

      Note that when it contains extra GET arguments, the base URL ends in an ampersand, **&**; if there are no extra arguments, then it ends in a question mark, **?**. In addition if there are any extra arguments in the base URL, then they must not be one of the standard service arguments defined below in this section.

      Every query to a given image query service uses the same base URL, and this is the URL which is registered as the "accessURL" for the service (see the "Registering a Compliant Service" section below).

   2. Constraints expressed in the standard "application/x-www-form-urlencoded" form, which is a list of ampersand-delimited GET arguments, each of the form:
      _<name>_=_<value>_

      > **Examples:**
      > ```
      > POS=180.567,-30.45&SIZE=0.0125
      > POS=22.567,61.45&SIZE=0.0125,0.0250&band=j
      > ```

      The constraints represent the query parameters which can vary for each successive query.

      The baseURL and constraint list are concatenated to form the query.

      > **Example:**
      > ```
      > http://myimages.org/cgi-bin/VOimq?POS=180.567,-30.45&SIZE=0.0125
      > ```

2. **Region of Interest.** The service **must** support the following two parameters which are used to specify the rectangular region of interest (ROI) on the sky to be used to search for images.
   **POS**
   The position of the region of interest, expressed as the right-ascension and declination of the field center, in decimal degrees using the ICRS coordinate system. A comma should delimit the two values; embedded whitespace is not permitted. Example: "POS=12.821,-33.4".
   **SIZE**

The coordinate angular size of the region given in decimal degrees. The region may be specified using either one or two values. If only one value is given it applies to both coordinate axes. If two values are given the first value is the angular width in degrees of the right-ascension axis of the region, and the second value is the angular width in degrees of the declination axis. Note that the angular width along the right-ascension axis is *not* the coordinate width of the region in right-ascension, which is equal to the coordinate angular width multiplied by cos(DEC).

A special case is SIZE=0. For an atlas or pointed image archive this tests whether the given point is in any image. For a cutout or mosaic service this will cause a default sized image to be returned. The default image size is service-defined and may be a value considered appropriate for the service, for the given image or data collection being accessed, or for the object (if any) at the given position.

These parameters define a nonrotated, rectangular region on the sky, having the specified angular extent in right ascension and declination, using the cartesian (CAR) projection with the region center (POS) as the reference point. The cartesian projection is used here as it is simple and can scale to the whole sky, and works about as well as anything else for small regions. ***Note:*** the use of the CAR projection to define the ROI has nothing to do with what projection we choose for any actual generated images.

---

**Technical Note**

```
Large query regions are well defined for the CAR projection [FITS WCS
Paper 2] used for the ROI. For example, a region with DECSZ=20 and
RASZ=360, with POS anywhere on the equator, defines a region 20 degrees
wide covering the entire celestial sphere at the equator. A region with
DECSZ=90 and RASZ=180 at 45 degrees N would cover one half of the northern
hemisphere. Regions which extend past the pole are defined (they reflect
about the pole) but are not likely to be very useful. The ROI is undefined
at the poles, but can be approximated well enough for the purposes of a
query by assuming RASZ=360.
```

---

In addition, the service **may** support the following query constraint which is used to pre-select images on the server side by specifying the type of intersection of a candidate image with the region of interest:

**INTERSECT**

A parameter that indicates how matched images should intersect the region of interest. The allowed values are:
- COVERS -- The candidate image covers or includes the entire ROI.
- ENCLOSED -- The candidate image is entirely enclosed by the ROI.
- CENTER -- The candidate image overlaps the center of the ROI.
- OVERLAPS -- The candidate image overlaps some part of the ROI.

If this parameter is not present, **INTERSECT=OVERLAPS** is assumed. Calculations need not be exact, e.g., a nonrotated bounding box approximation may be used to compute the type of intersection of a candidate image with the ROI. If the client requires a more precise measure, the spatial intersection a target image with the ROI can be computed precisely using the WCS metadata returned in the output VOtable. For a cutout or mosaicing service this parameter refers to the portion of the generated image containing valid (non-blank) data.

Note that case is important for these names: the service **must** permit these parameter names to be input in upper-case (the service may permit parameter names to be case insensitive since this permits upper-case names). By convention, upper case is used here for interface-defined terms in URLs, reserving lower case for service-defined parameters or parameter values.

3. **Image Generation Parameters.** The service **may** support the following image generation parameters (IGP) which are used to specify the desired size, scale, and orientation of the output image. These parameters are provided for the cutout and mosaicing services which generate images upon demand. Services which do not support these parameters may ignore them, but **must** allow them to be present without error; the client is allowed to pass these parameters to an image service whether it supports them or not. If the image generation parameters are input the ROI must still be specified, and the position thus defined **must** be consistent with the region (if any) defined by the IGP.

**NAXIS**

The size of the output image in pixels. This is a vector-valued quantity, expressed as

"NAXIS=<width>,<height>". If only one value is given it applies to both image axes. Default: determined from the ROI (see below). This is the only image generation parameter likely be supported by a cutout service.

**CFRAME**
The coordinate system reference frame, selected from ICRS, FK5, FK4, ECL, GAL, and SGAL (these abbreviations follow CDS Aladin). Default: ICRS.

**EQUINOX**
Epoch of the mean equator and equinox for the specified coordinate system reference frame (CFRAME). Not required for ICRS. Default: B1950 for FK4, otherwise J2000.

**CRPIX**
The coordinates of the reference pixel, expressed in the pixel coordinates of the output image, with [1,1] being the center of the first pixel of the first row of the image. This is a vector-valued quantity; if only one value is given it applies to both image axes. Default: the image center.

**CRVAL**
The world coordinates relative to CFRAME at the reference pixel. This is a vector-valued quantity; both array values are required. Default: the region center coordinates (POS) at the center of the image, transformed to the output coordinate system reference frame if other than ICRS. If CRPIX is specified to be other than the image center the corresponding CRVAL can be computed, but should be specified explicitly by the client.

**CDELT**
The scale of the output image in decimal degrees per pixel. A negative value implies an axis flip. Since the default image orientation is N up and E to the left, the default sign of CDELT is [-1,1]. This is a vector-valued quantity; if only one value is given it applies to both image axes, with the sign defaulting as specified above. Default: implied (see below), otherwise service-specific.

**ROTANG**
The rotation angle of the image in degrees relative to CFRAME (an image which is unrotated in one reference frame may be rotated in another). This is the rotation of the WCS declination or latitude axis with respect to the second axis of the image, measured in the counterclockwise direction (as for FITS WCS, which is in turn based on the old AIPS convention). Default: 0 (no rotation).

**PROJ**
The celestial projection of the output image expressed as a three-character code as for FITS WCS, e.g., "TAN", "SIN", "ARC", and so forth. Default: TAN.

> **Technical Note**
> The easiest way to understand the IGP is to consider a typical use case. The IGP are intentionally chosen to be consistent with FITS WCS (with some simplifications). A common use of an image mosaicing service is to generate an image to spatially match some reference image so that the two images can be easily compared. In this case the values of the IGP are equivalent to the WCS of the reference image, with minor formatting changes, making the IGP trivial for the client to input. A consistent POS can be computed by merely evaluating the reference WCS at the image center, and transforming the resultant coordinate to ICRS if necessary. These should be the values input in the image query to a mosaicing service.

**Image Geometry Defaults.** The default size and extent of the desired output image are determined from these parameters as follows: if a size or scale term is given explicitly this is the value used, otherwise a default is computed based on the angular extent of the query region (SIZE). For example, if NAXIS$n$ and CDELT are given then the angular size of the image is determined by these parameters. If the NAXIS$n$ are given then the default image scale is determined by SIZE. If only CDELT is given then the image size in pixels is determined by the angular extent of the query region (SIZE). If only the region angular extent SIZE is given then the size (in pixels) and scale of the image is determined by the service to best fit the data and the region of interest.

In the simplest case, all image generation parameters may be omitted, and the generated output image defaults to the position and size of the query region. If any image generation parameters are specified these override the ROI-implied defaults.

4. **Image Format.** The service **must** support a parameter with the name **FORMAT** to indicate the desired format or formats of the images referenced by the output table. The value is a comma-delimited list where each element can be any recognized MIME-type. Typically, these will be of the major type `image`; however, "`text/html`" can be specified to request URLs for HTML documents describing and/or previewing the image.

**Single-Format Examples:**

```
    image/fits
    image/png
    image/jpeg
    text/html
```

In addition, these special values are defined:

- ○ "ALL" -- Denotes all formats supported by the service.
- ○ "GRAPHIC" -- Denotes any of the following graphics formats: JPEG, PNG, GIF. These are types typically supported "in-line" by web-browsers.
- ○ "METADATA" -- Denotes a *Metadata Query*: no images are requested; only metadata should be returned. This feature is described in more detail in [section 6.1](#).

The following optional syntax is defined for "GRAPHIC":

- ○ "GRAPHIC-ALL" -- The output table should list every graphics format available for each image. For services that support multiple graphics formats this can result in a large output table.
- ○ "GRAPHIC-jpeg,png,gif" -- This specifies the client preference for graphics image format. In the example, "jpeg" is the first choice, and "gif" is the last choice, but the client will accept any graphics format listed. The service will return the first such format listed which it supports. If this syntax is used it must be the last element in the FORMAT list.

If only FORMAT=GRAPHIC is specified, the default is to return one graphics format entry for each image in the output table, with the format being selected by the service. Clients should only use this shorthand version if they can accept any of the three standard graphics formats.

> **Format List Example:**
> `image/fits,text/html,GRAPHIC-jpeg,png`

The example above specifies the desired output image formats to be image/fits, text/html, and one of image/jpeg or image/png. Any formats matching this list are returned for each image.

No requirements are placed on compliant services on the image formats that they must support; however, if images are supported, their MIME major type must be `image`.

If formats indicated by this parameter are not supported by the implementing service, the returned table should be empty. If the format is not specified **FORMAT=ALL** is assumed.

5. **Table Verbosity.** The service **may** support an optional parameter with the name **VERB** (denoting "verbose") whose value is a nonnegative integer. This parameter indicates the desired level of information to be returned in the output table, particularly the number of columns to be returned to describe each image. The following guidelines are recommended for determining which columns should be output at different verbosity levels:
   - ○ 0 -- The output table should contain only the minimum columns required by [section 4.2](#).
   - ○ 1 -- In addition to level 0, the output table should contain columns sufficient for uniquely describing the image.
   - ○ 2 -- In addition to level 1, the output table should contain, if possible, columns that contain values for all parameters supported as query constraints.
   - ○ 3 -- The output table should return as much information about the images as possible. A table metadata query automatically implies the highest level of verbosity.

   Services that do not support this parameter **must** permit it to be present without error.

6. **Service-Defined Parameters.** The service **may** support additional service-specific parameters. The names, meanings, and allowed values are defined by the service. The names need not be upper-case; however, they should not match any of the reserved parameter names defined above. Values must be simple and describable through the output of the **metadata query** as described below.

## 4.2 Successful Output

The output returned by an **Image Query** is a [VOTable](#), an XML table format, returned with a MIME-type of `text/xml;content=x-votable`. The table lists all the images available to the client that match the query constraints. The following requirements are placed on the contents of the table when the query successfully returns a list of images:

1. The VOTable **must** contain a RESOURCE element, identified with the tag `type="results"`,

containing a single TABLE element which contains the results of the query. The VOTable is permitted to contain additional RESOURCE elements, but the usage of any such elements is not defined here. If multiple resources are present it is recommended that the query results be returned in the first resource element.

2. The RESOURCE element **should** contain an INFO with `name="QUERY_STATUS"`. Its value attribute should set to "OK" if the query executed successfully, regardless of whether any matching images were found. All other possible values for the value attribute are described in <u>section 4.3</u> below.

> **Examples:**
> ```
>     <INFO name="QUERY_STATUS" value="OK"/>
>
>     <INFO name="QUERY_STATUS" value="OK"> Successful Search</INFO>
> ```

3. Each table row represents a different image available to the client.
4. Each row of the output VOTable **must** contain FIELDs where the following <u>UCDs</u> have been set. These attributes apply to graphics images as well as to FITS images, and are required to associate standard image metadata (such as a WCS) with images of all types, and to be able to return image metadata separately from the image itself.

### Image Metadata

- Exactly one field **must** have `ucd="VOX:Image_Title"`, with `datatype="char"`, and `arraysize="*"`, containing a short (usually one line) description of the image. This should concisely describe the image to a user, typically identifying the image source (e.g., survey name), object name or field coordinates, bandpass/filter, and so forth. Note that a client application may present the user with a combined list of images from many different queries and sources: the title string is important to identify each image to the user.
- Exactly one field **should** have `ucd="INST_ID"`, with `datatype="char"`, and `arraysize="*"`, identifying the instrument or instruments used to make the observation, e.g., `STScI.HST.WFPC2`.
- Exactly one field **should** have `ucd="VOX:Image_MJDateObs"`, with `datatype="double"`, representing the mean modified Julian date of the observation. By "mean" we mean the midpoint of the observation in terms of normalized exposure times: this is the "characteristic observation time" and is independent of observation duration.
- Exactly one field **must** have `ucd="POS_EQ_RA_MAIN"`, with `datatype="double"`, representing the ICRS right-ascension of the center of the image.
- Exactly one field **must** have `ucd="POS_EQ_DEC_MAIN"`, with `datatype="double"`, representing the ICRS declination of the center of the image.
- Exactly one field **must** have `ucd="VOX:Image_Naxes"`, with `datatype="int"`, specifying the number of image axes.
- Exactly one field **must** have `ucd="VOX:Image_Naxis"`, with `datatype="int"`, and `arraysize="*"`, with the array value giving the length in pixels of each image axis.
- Exactly one field **must** have `ucd="VOX:Image_Scale"`, with `datatype="double"`, and `arraysize="*"`, with the array value giving the scale in degrees per pixel of each image axis.
- Exactly one field **must** have `ucd="VOX:Image_Format"`, with `datatype="char"`, and `arraysize="*"`, specifying the MIME-type of the object associated with the image acref, e.g., "image/fits", "text/html", and so forth.

> **Note:**
> UCDs starting with "VOX:" (as in "VO-experimental") are not standard UCDs as defined by the <u>IVOA UCD Standard</u>. These terms are data model attributes as much as metadata elements and a framework for integrating the two had not been developed at the time of writing. Thus, the prefix, "VOX:", defines a temporary namespace for UCD-like terms defining experimental concepts used in our VO prototypes. Existing UCDs are shown in all upper case without the "VOX:" namespace identifer.

### Coordinate System Metadata

This defines a basic sky projection world coordinate system (WCS) for the image. Each image description **should** include coordinate system metadata, but this is not required. The WCS metadata included here is FITS WCS-like but has been simplified: if the image contains an actual FITS WCS this may contain additional information not provided here.

- Exactly one field **should** have `ucd="VOX:STC_CoordRefFrame"`, with `datatype="char"`, and `arraysize="*"`, representing the coordinate system reference frame, selected from "ICRS", "FK5", "FK4", "ECL", "GAL", and "SGAL".
- Exactly one field **may** have `ucd="VOX:STC_CoordEquinox"`, with `datatype="double"`,

representing the Equinox (not required for ICRS) of the coordinate system used for the image world coordinate system (WCS). This should match whatever is in the image WCS and may differ from the default ICRS coordinates used elsewhere.

○ Exactly one field **should** have ucd=`"VOX:WCS_CoordProjection"`, with `datatype="char"`, and `arraysize="3"`, with the array value being the three-character code ("TAN", "ARC", "SIN", and so forth) specifying the celestial projection, as for FITS WCS.

○ Exactly one field **should** have ucd=`"VOX:WCS_CoordRefPixel"`, with `datatype="double"`, and `arraysize="*"`, with the array value specifying the image pixel coordinates of the WCS reference pixel. This is identical to "CRPIX" in FITS WCS.

○ Exactly one field **should** have ucd=`"VOX:WCS_CoordRefValue"`, with `datatype="double"`, and `arraysize="*"`, with the array value specifying the world coordinates of the WCS reference pixel. This is identical to "CRVAL" in FITS WCS.

○ Exactly one field **should** have ucd=`"VOX:WCS_CDMatrix"`, with `datatype="double"`, and `arraysize="*"`, with the array (matrix) value specifying the WCS CD matrix. This is identical to the "CD" term in FITS WCS, and defines the scale and rotation (among other things) of the image. Matrix elements should be ordered as CD[i,j] = [1,1], [1,2], [2,1], [2,2].

The above terms are based on the following two proposals: Space-Time Coordinate Specification for VO Metadata ("STC" above), and Representations of World Coordinates in FITS ("WCS" above).

If the coordinate system metadata defined here is omitted, the client can estimate the WCS of the image from the required image metadata. That is, the reference pixel defaults to the image center, the reference value is given by the RA and DEC values at the center of the image, the coordinate system is ICRS, and the image is assumed to be unrotated with no flipped axes (N up and E to the left), with the given scale. This estimated WCS is not sufficient to define the exact image footprint or where the image pixels are on the sky, but it may be sufficient for some purposes such as selecting images for retrieval. A standard WCS should still be returned in the returned image if at all possible.

**Spectral Bandpass Metadata**

This defines a simple model to characterize the spectral bandpass of the image. The image description **should** include spectral bandpass metadata for the image, but this is not required (it is highly desirable to permit automated multiwavelength analysis). The simplified model presented here is based on an analysis by Jonathan McDowell (CXC/SAO).

○ Exactly one field **should** have ucd=`"VOX:BandPass_ID"`, with `datatype="char"`, and `arraysize="*"`, identifying the bandpass by name (e.g., "V", "SDSS_U", "K", "K-Band", etc.).

○ Exactly one field **should** have ucd=`"VOX:BandPass_Unit"`, with `datatype="char"`, and `arraysize="*"`, identifying the units used to represent spectral values, selected from "meters", "hertz", and "keV". No other units are permitted here; the client application may of course present a wider range of units in the user interface.

○ Exactly one field **should** have ucd=`"VOX:BandPass_RefValue"`, with `datatype="double"`, specifying the characteristic (reference) frequency, wavelength, or energy for the bandpass model.

○ Exactly one field **should** have ucd=`"VOX:BandPass_HiLimit"`, with `datatype="double"`, specifying the upper limit of the bandpass.

○ Exactly one field **should** have ucd=`"VOX:BandPass_LoLimit"`, with `datatype="double"`, specifying the lower limit of the bandpass.

The characteristic value and upper and lower limits should be chosen to best model the actual bandpass of the image; exactly how this is done is up to the data provider. Typical values would be the mode or central wavelength/frequency of the transmission function for the reference value, with the lower and upper cutoffs chosen to include essentially all (not necessary 100%) of the transmitted energy. At this point there is no provision for modeling the actual transmission function. Our intent here is merely to model the spectral bandpass of the image in the local rest frame (as for a filter), not to provide a full spectrophotometric calibration for the image.

**Processing Metadata**

Processing metadata is used to describe any processing done to the data as presented by the image service. This is especially important if the service changes the data in any way, for example a mosaicing service which reprojects image data which has already been sampled into a pixel grid.

○ Exactly one field **should** have ucd=`"VOX:Image_PixFlags"`, with `datatype="char"`, and

`arraysize="*"`, specifying the type of processing done by the image service to produce an output image pixel. The string value should be formed from some combination of the following character codes:

- C -- The image pixels were copied from a source image without change, as when an atlas image or cutout is returned.
- F -- The image pixels were computed by resampling an existing image, e.g., to rescale or reproject the data, and were filtered by an interpolator.
- X -- The image pixels were computed by the service directly from a primary data set hence were not filtered by an interpolator.
- Z -- The image pixels contain valid flux (intensity) values, e.g., if the pixels were resampled a flux-preserving interpolator was used.
- V -- The image pixels contain some unspecified visualization of the data, hence are suitable for display but not for numerical analysis.

For example, a typical image cutout service would have `PixFlags="C"`, whereas a mosaicing service operating on precomputed images might have `PixFlags="FZ"`. A preview page, graphics image, or a pixel mask might have `PixFlags="V"`. An image produced by sampling and reprojecting a high energy event list might have `PixFlags="X"`. If not specified, `PixFlags="C"` is assumed.

## Access Metadata

Access metadata includes any information describing either the data or the service which is needed to access the data.

- Exactly one field **must** have ucd=**"VOX:Image_AccessReference"**, with `datatype="char"` and `arraysize="*"`, specifying the URL to be used to access or retrieve the image. Since the URL may contain XML metacharacters the URL is can be enclosed in an XML CDATA section (`<![CDATA[...]]>`) or otherwise URL encoded (see URI Specification) to escape any embedded metacharacters.

  It is possible and permissible for an image query to reference an image which cannot be accessed, e.g., because the image metadata is available online but the image itself is not. In this case the access reference value should be given as "NONE", without a CDATA enclosure.

  Since SIA is a URL-based interface the access reference is a simple URL. If the client issues a HTTP GET request using this URL, and the request is successful, the client will receive a VOX:Image_Format document in return. In most cases this will be an image with a MIME-type such as "image/fits", but if the image reference points to a preview page, the document type could be "text/html". ***Note:*** the access reference URL does not stage the data and return another image reference, it returns the image itself as the response to the GET.

  The access reference URL need not point to an actual existing online image. In many cases the URL will reference a virtual image, which will be created on-the-fly by the image service when accessed.

  Image access references (acrefs) are strings which uniquely identify a (possibly virtual) image within the global name space of the Web. Acref strings are used like file pathnames to tag or index images, hence should be static in nature (meaning the string should be computed once by the service and thereafter merely copied about). Acrefs can be passed about among cooperating hosts in a distributed computing scenario, hence should access the same image regardless of the address of the client. An acref is a *runtime* image reference, generated by the image service at query time, and may be invalid after an interval of time (e.g., hours or days), as defined by the service.

- Exactly one field **may** have ucd=**"VOX:Image_AccessRefTTL"**, with `datatype="int"`, specifying the minimum time to live in seconds of the access reference.
- Exactly one field **should** have ucd=**"VOX:Image_FileSize"**, with `datatype="int"`, representing the *actual or estimated size of the encoded image in bytes* (not pixels!). This is useful for image selection and for optimizing distributed computations.

## Resource Metadata

- The table **may** contain other service-specific columns not specified above. Tagging of columns via the FIELD's UCD attribute is strongly encouraged where UCDs match the column values. If a service-specific column contains values associated with a query input parameter, it must be described according to the specification for table metadata query response ([section 4.1.5](#)). If the column is not associated with an input parameter, such a description is not required but is encouraged.

5. **Constant Valued FIELDs.** If a FIELD of the table will have the same value for every row of the table it is permissible to represent the field as a PARAM of the resource rather than a table FIELD. Both FIELDs and PARAMs can be used to define image metadata: a FIELD allows a different value to be specified for every image, whereas PARAM allows metadata to be defined globally for all images. The same attribute tags (UCD, datatype, arraysize, type="trigger", etc.) must be defined in either case.

## 4.3 Error Responses and Other Unsuccessful Results

An INFO element within the "results" RESOURCE element of the the VOTable is used to indicate success or failure of the image query. As described in the previous section, the INFO element must have `name="QUERY_STATUS"`; if the query is successful (regardless of whether any image rows are returned) the value attribute is set to "OK". The remainder of this section defines other possible values to indicate that the query was unsuccessful in some way.

When the query is unsuccessful, the contents of INFO element (i.e. its PCDATA child node) **should** contain an error message suitable for display. Implementation of this feature is encouraged but not required.

---

**Note:**
    The encoding of errors defined here differs substantially from the [Simple Cone Search specification](#). The Cone Search specifies the use of a PARAM with an error message in the value attribute. This specification uses controlled vocabulary for the value, and places the optional error message in the element's contents.

---

When the query is unsuccessful (in any of senses described below), the resulting VOTable is not required to contain any other elements as specified in the [previous section](#); although, it is not an error to do so.

The other allowed values for value attribute besides "OK" are:

1. **ERROR**
   The server failed to process the query. Possible reasons include:

   - The input query contained a syntax error.
   - The way the query was posed was invalid for some reason, e.g., due to an invalid query region specification.
   - A constraint parameter value was given an illegal value; e.g. `DEC=91`.
   - The server trapped an internal error (e.g., failed to connect to its database) preventing further processing.

   In this case, the inclusion of a descriptive error message is strongly encouraged.

2. **OVERFLOW**
   The query produced results that exceeded the limits of the service in some way. This may be in the size of the images requested or in the number of matching images. In this case, the service **should** include an error message indicating the nature of the overflow condition.

---

**Examples:**
```
<INFO name="QUERY_STATUS" value="ERROR">DEC out of range: DEC=91</INFO>
<INFO name="QUERY_STATUS" value="ERROR">SZ given in addition to RASZ and/or
DECSZ</INFO>
<INFO name="QUERY_STATUS" value="OVERFLOW">Number of matching images exceeds limit
of 100</INFO>
<INFO name="QUERY_STATUS" value="OVERFLOW">Requested image size exceeds limit of 1
degree on a side</INFO>
```

---

## 5. Image Retrieval

The image retrieval request (**getImage** web method) allows a client to retrieve a single image given an access reference or "acref" as returned by a prior image query. In the case of SIA the acref is a simple URL since SIA is URL-based. Since an image query is required to obtain an acref, no requirements are placed on the form the acref takes. This has the effect of hiding the details of the acref URL from the client, making it easy to layer an implementation of the getImage web method on top of an existing image retrieval service, and making it easier to hide changes to the implementation of existing services.

## 5.1 Input

The input to the getImage web method is the image acref for the indicated image or virtual image. The acref for a particular image is obtained through a prior call to the **Image Query** web method.

## 5.2 Successful Output

The output of getImage **must** be a single image or image document returned with a MIME-type identifying the file format. If an image is returned it must conform to the simple image data model as outlined in Image Service Types above. When the input acref points to a physical image the primary type of the MIME code should be "image/". Other MIME types are allowed, depending on the capabilities of the image service; for example, a MIME-type of "text/html" may be used when the acref URL points to an HTML description and/or preview of the image.

If a FITS image is returned the image **should** contain a valid **FITS WCS**. Any areas of the image which do not contain valid data, e.g., because the requested region extends beyond the bounds of the source image, **should** be flagged with a blank value, using the FITS keyword **BLANK** to identify the blank fill value used.

## 5.3 Error Response

If a condition is encountered that prevents the requested image from being downloaded, the output **must** be a VOTable with a single RESOURCE element containing an INFO child with `name="QUERY_STATUS"`. The allowed values for this INFO are the same as those defined for the Image Query; in addition, the following additional attribute **may** be supported:

**DEFERRED**
This indicates that the requested image is not yet available for some reason but that it will be at some time in the future. Clients that receive this type of message can periodically try (poll) the given acref URL until the image becomes available.

# 6. Service Metadata

Compliant image services describe themselves in two ways. They advertise their availability by registering with a registry service, including supplying service metadata to characterize the service and any associated data collections. They also advertise their capabilities through support of the special metadata query signaled through the FORMAT=METADATA parameter (cf. section 4.1.4). When the service provider registers an SIA service, the registry can execute the metadata query to collect the capability metadata. The gathering of service and capability metadata from all SIA services enables a client to use the registry to discover the image services most appropriate for a particular computation.

## 6.1. Metadata Query

A compliant service **must** support image queries with FORMAT=METADATA, used to query the service metadata; only metadata is returned by the service in this case. In particular, the response to this query advertises two things about the service:

* which input parameters the service supports (see section 4.1).
* the columns it will return.

Note that the SIA specification is designed so that a client does not need know this information to make use of the service. The metadata query is mainly intended for use by a central registry that will collect this information so that users can search for services by their capabilities. It is also useful for comminicating non-standard input parameters.

The overall structure of the VOTable by a metadata request is the same as described in section 4.2 except that it contains no table rows. (In fact, it can in practice look exactly like a normal image query

response containing no matching images.)

Each input parameter supported by the service should be listed as a PARAM element of the RESOURCE that normally contains the image table. Each PARAM should have a `name` attribute of the form "`INPUT:`*param_name*", where *param_name* is the parameter name as it should appear in the query URL. For example, `name=`'`INPUT:POS`' refers to the "POS" input parameter. All input parameters meant to be available to clients of the service must be listed as PARAM elements, including required parameters (POS, SIZE, and FORMAT), optional parameters (defined in section 4.1), and non-standard parameters specific to the service. The PARAM may contain a `value` attribute which should contain the default value that will be assumed if the parameter is not set in the query input URL. Implementors are encouraged to include, as children of the PARAMs, DESCRIPTION elements to describe the parameter and (where appropriate) VALUES elements to given allowed ranges or values.

---

**Example:** The input parameter listing below from a <u>Pointed Image Archive</u> shows that in addition to supporting the required parameters (POS, SIZE, and FORMAT), it also supports the optional parameters INTERSECT and VERB as well as its own non-standard parameter, "telescope".

```
<RESOURCE type="results">
  <DESCRIPTION>ADIL Simple Image Access Service</DESCRIPTION>
  <INFO name="QUERY_STATUS" value="OK"/>
  <PARAM name="INPUT:POS" value="0,0" datatype="double">
    <DESCRIPTION>Search Position in the form "ra,dec" where
                 ra and dec are given in decimal degrees in the ICRS
                 coordinate system.</DESCRIPTION>
  </PARAM>
  <PARAM name="INPUT:SIZE" value="0.05"  datatype="double">
    <DESCRIPTION>Size of search region in the RA and Dec. directions</DESCRIPTION>
  </PARAM>
  <PARAM name="INPUT:FORMAT" value="ALL" datatype="char" arraysize="*">
    <DESCRIPTION>Requested format of images</DESCRIPTION>
    <VALUES>
      <OPTION value="image/fits"/>
      <OPTION value="image/gif"/>
      <OPTION value="text/html"/>
      <OPTION value="ALL"/>
      <OPTION value="GRAPHIC"/>
      <OPTION value="METADATA"/>
    </VALUES>
  </PARAM>
  <PARAM name="INPUT:INTERSECT" value="OVERLAPS" datatype="char" arraysize="*">
    <DESCRIPTION>Choice of overlap with requested region</DESCRIPTION>
  </PARAM>
  <PARAM name="INPUT:VERB" value="1" datatype="int">
    <DESCRIPTION>Verbosity level, controlling the number of columns returned</DESCRIPTION>
  </PARAM>
  <PARAM name="INPUT:telescope" datatype="char" arraysize="*" value="HST">
    <DESCRIPTION>Telescope name</DESCRIPTION>
  </PARAM>
</RESOURCE>
...
```

---

The columns that are returned by an image query are described with the FIELD elements exactly as they are described in a normal image query response (section 4.2).

When FORMAT=METADATA is given, all other input parameters are normally ignored. An exception is the VERB parameter: if given, the service **may** choose to describe only those columns that will be returned for that verbosity level. If the service chooses to support the VERB parameter in this way, it must provide a PARAM with `name="VERB"` and a `value` attribute giving the actual verbosity level being described.

---

**Example:** The following PARAM could appear after the PARAM elements shown in the previous example:

```
<PARAM name="VERB" value="1" datatype="int" />
```

---

If the service chooses to ignore the VERB parameter, than it should describe all of the columns returned at the highest verbosity level supported.

## 6.2. Registering a Compliant Service

In order to advertise an SIA service it should be registered with an IVOA compliant registry service. The general metadata that are associated with the service are drawn from the set in "Resource and Service Metadata for the Virtual Observatory" (which in turn draws from the Dublin Core). These metadata are made concrete by expressing as XML schema as described in "VOResource: an XML Encoding Schema for Resource Metadata". This latter document describes how individual protocols may define extensions to the core schema to define protocol specific metadata.

The following additonal service specific metadata are required for registration (with the metadata that are defined by the "Resource and Service Metadata for the Virtual Observatory" document marked with an [R]) :

- **Type.ImageService** [R]: the category of Image Service that this services falls into. The allowed values (defined in section 3) are:
  - **Cutout**
  - **Mosaic**
  - **Atlas**
  - **Pointed**
- **Coverage.Spectral** [R]: The spectral wavebands covered by this image service given by one or more of the following: **gamma, xray, euv, uv, optical, infrared, microwave,** and **radio**.

  > **Note:**
  >     The bandpass associated with these wavebands are not strictly defined.

- **Coverage.Temporal** [R]: The approximate time range that is covered by the image data.

  > **Note:**
  >     The format of the value is determined by STC.

- **Coverage.Spatial** [R]: The approximate area or areas of the sky covered by the image data.

  > **Note:**
  >     The format of the value is determined by STC.

- **MaxQueryRegionSize**: The maximum image query region size, expressed in decimal degrees. A value of 360 degrees indicates that there is no limit and the entire data collection (entire sky) can be queried. The service should return OVERFLOW if this value is exceeded.
- **MaxImageExtent**: The maximum image extent on the sky in decimal degrees, e.g., "1.0 1.0".
- **MaxImageSize**: The maximum image size in pixels, e.g., "8192 8192".
- **MaxImageFileSize**: The maximum image file size in bytes, e.g., "300m". The suffixes "k" (kilobytes), "m" (megabytes), and "g" (gigabytes) are permitted.
- **MaxRecords**: The largest number of records that the **Image Query** web method will return. The service should return OVERFLOW if the maximum number of records is exceeded.
- **ImageFormats**: The MIME types of the image formats which the service can return.
- **ImageQueryBaseURL**: The base URL clients should use for the **Image Query** web method (see section 4.1).

The mapping between these metadata names and the element names in the V1.0 registry schema is given in appendix B.

# 7. Usage Examples

## 7.1 Client Examples

### Simple Query

In this example the client requests a service to list all the image data available for some region. Depending upon how the image service being queried was set up it may access one or more surveys or other image data collections. For our example we query a region centered at RA=10.63 and DEC=41.25 (ICRS) with an extent of 0.15 degrees of arc in each axis. The image query request is as follows:

```
http://archive.foo.edu/vo/sim/queryImages?POS=10.63,41.25&SIZE=0.15
```

This query will work for any image service. For an atlas or pointed image service the service will list any image which intersects the region, regardless of the size, orientation, resolution, bandpass, etc., of the image. For a cutout or mosaicing service only images equal in size to the referenced region would be listed (since these services will generate images of the requested size), but they could still have any orientation, scale, resolution, bandpass, and so forth.

It is up to the client to examine the returned VOTable to determine which if any of the images described are to be downloaded. The image metadata returned in the table is used to perform image selection. For example, the WCS information in the table may be used to select only images which sufficiently overlap or match the region the client is interested in. The Image_FileSize attribute may be used to reject images which are too large to be worthwhile downloading. Any technique could be used for image selection, e.g., selection could be automated based on the returned metadata, or the image metadata (specifically the WCS) could be used to generate a tagged graphics marker in an image display which the user could interactively select to download the associated image.

Although SIA provides some facilities for server-side image selection, e.g., the optional INTERSECT parameter, ultimately the algorithm used for image selection has to be controlled by the client since there are potentially infinitely many client applications, and only the client knows what it is trying to do. Hence the image metadata returned as the query response plays a central role in image access.

Once the images to be retrieved have been selected the individual images can be fetched using the image access reference returned in the table record referencing the image (field Image_AccessReference). The string value of this field is the URL of the "image" (real or virtual) to be retrieved. This field is opaque to the client, but an example might be as follows:

```
http://archive.foo.edu/vo/sim/getImage?survey=dposs&filter=V&pos=10.63,41.25&size=0.15
```

## Survey Cutout

An image service which is set up to return cutouts from a single survey is no different than the simple query above, except that the service is dedicated to a single survey. In this case the image query isn't really doing a search at all, rather it is used merely to generate the access reference required to retrieve the desired image cutout. The returned VOTable will list one or more images depending upon whether the desired band (filter) was specified in the query. The image referenced by each access reference URL is then retrieved and the survey cutout is complete. An attempt to access (query) a region outside the coverage of the survey will result in a no data being returned.

The image service registry is used to discover all services available for the desired survey. If the same resource is available from multiple services these can be used for load distribution, to select the maximum network bandwidth between two endpoints, to fail-over in the event that a given service is offline, and so forth.

## Distributed Query

A typical distributed query would access multiple services to gather data on a region or regions from multiple wavelength regimes. Typically the image services to be accessed are widely distributed.

The registry is first used to select the list of image services to be queried. Assume we are looking for all the data available on a given region. The same query (as in Simple Query, above) would then be posed to all the target services. The query results would be analyzed for each service and the selected images retrieved. Some form of analysis would then be performed on the resultant set of images.

This is an inherently parallel operation, with the same operations being performed independently for each image service. Hence all queries can proceed concurrently - in principle the distributed operation can be completed in not much more time than it would take to query a single service.

## Image Generation

To illustrate how an image generation service is used we first obtain a reference image, and then try to use a mosaicing service to generate an image which spatially matches our reference image.

For our example we have selected as our reference image a 256x256 pixel image cutout from the NDWFS survey. This is a wide field optical CCD mosaic survey: the data has been heavily processed, with many frames being dithered and stacked to produce the final online image data. Since the data had to be resampled it has been reprojected to a purely analytic, unrotated projection, TAN in this case. The entire 3 degree-wide field shares the same projection, hence cutouts are characterized by a reference pixel located well outside of the image. A portion of the FITS header of this reference image follows (the OBJECT keyword shown here is a good example of an Image_Title string by the way).

```
OBJECT = 'NDWFS J142859.86+353716.4 Bw-band'
EQUINOX = 2000. / Equinox of coordinate system
```

```
RADECSYS = 'ICRS ' / Default coordinate system
DATE-OBS= '1999-04-12' / UT Date of Earliest Obs. Used In Combined Image
FILTER = 'Bw NDWFS k1025' / Filter name(s)
CTYPE1 = 'RA---TAN' / Coordinate type
CTYPE2 = 'DEC--TAN' / Coordinate type
CRVAL1 = 218.0238 / Coordinate reference value
CRVAL2 = 34.27986 / Coordinate reference value
CRPIX1 = -8658. / Coordinate reference pixel
CRPIX2 = -18627. / Coordinate reference pixel
CD1_1 = -7.1666666666667E-5 / Coordinate matrix
CD2_1 = 4.3883176969447E-21 / Coordinate matrix
CD1_2 = 4.3883176969450E-21 / Coordinate matrix
CD2_2 = 7.16666666666673E-5 / Coordinate matrix
```

If we take our simple query example and add image generation parameters to match our reference image the enhanced query would look as follows:

```
http://archive.foo.edu/vo/sim/queryImages?POS=217.2494167,35.6212417&SIZE=0.0183333&
NAXIS=256&CFRAME=FK5&CRPIX=-8658,-18627&CRVAL=218.0238,34.27986&CDELT=7.166667E-5
```

Here we have omitted those image generation parameters which agree with the defaults, i.e., EQUINOX (J2000), PROJ (TAN) and ROTANG (0.0). Since the image is unrotated CDELT is given by the diagonal terms of the CD matrix, CD1_1 and CD1_2. The off-diagonal terms are effectively zero in the example (4.388E-21), to within the precision of the WCS. POS was computed by evaluating the WCS of the reference image at the image center, X=Y=128.5. A more general procedure for calculating the image scale and rotation terms from the CD matrix is given in WCS Paper II.

If we submit this query to a mosaicing service the service will describe images which should exactly duplicate the spatial characteristics of the reference image. Other services will offer images which represent the best match to the requested image, in most cases using only the POS and SIZE parameters.

## 7.2 Service Examples

### Atlas or Pointed Image Service

An atlas or pointed image service serves a set of precomputed, fixed sized images from one or more surveys (atlas image service) or instruments (pointed image service). The basic task of an atlas or pointed image service is simple: process the image query URL, and return a VOTable listing all images which intersect the region of interest specified by the client. Each image entry will include an access reference URL which can be used by the client to fetch the image. If no images are found, the query is successful, but a VOTable containing no data rows (matched images) is returned.

The only parameters which this service must implement are POS, SIZE and FORMAT. Optional parameters the service may wish to implement include INTERSECT and VERB. Other useful service-defined parameters might include a parameter named "band" or "filter" of type BandPass_ID, as defined in the spectral bandpass section of Query Output above. Atlas or pointed image services should ignore the image generation parameters: all input parameters should however be allowed to be present in the input without error.

In operation the service should parse the image query URL and scan the set of images supported by the service to see which if any 1) match one of the requested image formats, and 2) intersect the query region. For each such image a record is formatted and added to the output VOTable: this is often the hardest part of the service to implement, as access to image metadata such as the WCS and spectral bandpass is required to fully describe the image. If the data is available online, an access reference URL (acref) should be computed which the client can invoke to retrieve the image.

For many services image access will be trivial, e.g., the acref could be a simple HTTP or FTP URL referencing the online image file. In other cases the URL might need to point to a CGI, web service, or other program which extracts the referenced image and returns it to the client. This is necessary if the image is stored in a form which violates the simple image data model. For example, CCD Mosaic frames stored as extensions in a FITS MEF must be extracted as separate images at access time: the extracted image can be returned directly to the client without staging it to disk. SDSS (as one example) stores atlas images of faint objects in a proprietary format which includes cutouts for each filter used by the survey. As each band of the atlas cutout is accessed it will need to be extracted and returned to the client separately. In cases like these where some server-side image extraction is needed it might prove worthwhile to cache the archived or extracted data on disk.

For an atlas or pointed image service `MaxQueryRegionSize` can be anything up to the entire sky (360

degrees). The `MaxImageExtent, MaxImageSize,` and `MaxImageFileSize` attributes should reflect the maximum size image the service can return; for an atlas or pointed image service this is usually the largest image available.

### Image Cutout Service

An image cutout service is very similar in operation to an atlas or pointed image service except that, instead of "returning" (describing) entire static images, the service extracts or "cuts out" a subraster from each image or survey field which intersects the query region. The position and size of the returned image cutout will match the POS and SIZE parameters of the region requested by the client. If no data is available for part of the cutout then the full cutout region is still returned, but with undefined regions of the image set to a blank value.

Cutout services often serve only a single uniform survey, in which case a query request will usually return only zero, one, or several images (one for each band or format at the given position). However this is not necessarily the case. If a cutout service serves multiple image data collections a query response may reference any number of images, just as for an atlas or pointed image service.

The required input parameters for a cutout service are the same as for an atlas or pointed image service, i.e., POS, SIZE, and FORMAT. INTERSECT is optional, but can be used to reject cutouts which do not contain sufficient valid data (e.g., INTERSECT=CENTER would require that the center of the cutout contain valid, or non-blank, data). Other service-defined parameters such as a Band_ID parameter are just as useful for a cutout service as for other types of image services.

In addition to the usual input parameters a cutout service will often support the NAXIS image generation parameter. This is desirable to allow the client to precisely control the size of the generated image in pixels. If NAXIS is specified, then assuming a fixed image scale, NAXIS overrides the default image size implied by the query region parameter SIZE.

For a cutout image service the query region size defines the default size of the image cutout, hence `MaxQueryRegionSize` should reflect the maximum size image cutout which can be generated. The `MaxImageExtent, MaxImageSize,` and `MaxImageFileSize` attributes should also reflect the maximum size image the service can return.

For a cutout service `MaxQueryRegionSize` and `MaxImageExtent` are probably identical. One can imagine allowing the query region size to be larger than the largest possible cutout, with the service proposing to tile the larger region with smaller cutouts. This is too complex to be worthwhile; if any tiling of large regions is to be done it is preferable to leave this up to the client since there are many ways to tile a large region.

Since cutout services create new images the service will need to generate a new image header as well as subset the image pixel array (at least for FITS format images). In particular the cutout service should propagate valid WCS data to the generated image cutout.

# Appendix A: Changes from Previous Versions

1. Removed section on "Staging" of results, as the suggested implementation was not complete.
2. Removed the inconsistency in the description of the SIZE parameter with respect to the CAR projection.
3. The Service Metadata section has been altered to be compliant with the http://www.ivoa.net /Documents/latest/VOResource.html and current practice with the 1.0 WD SIA schema extension. An appendix has been added with a more extensive description of how the metadata terms map to the 1.0 generation of registry schema.
4. Corrected VOTable examples to be valid VOTable 1.0.
5. General addition of references.
6. Removal of text which descibed what might be done to develop the protocol in "the future".

# Appendix B: Schema Extension for Simple Image Access Services

Resources are described using the VOResource schema, v1.0 [VOR] and its family of extensions. Simple Image Access services are specifically described using the VODataService, v1.0, and SIA, v1.0, extension schemas. Coverage information is encoded using the Space-Time Coordinates (STC) schema [STC]. These schemas, as well as others defined by the IVOA, may be found at http://www.ivoa.net/xml/

An SIA service is registered as resource of type vs:DataService with a vr:Capability of type sia:SimpleImageAccess, where vs:, vr: and sia: are namespace prefixes for the VODataService,

[VOResource](#) and [SIA](#) schemas respectively. in particular it is the sia:SimpleImageAccess capability with it's standardID="ivo://ivoa.net/std/SIA" attribute that marks the service as being a Simple Image Access Protocol service.

The following table enumerates the mapping of resource profile metadata defined in this specification with those defined in the XML schemas.

| SIA Metadatum | VOResource Metadatum | |
|---|---|---|
| | Schema Name | XPath Name |
| **Coverage.Spectral** | VODataService | `coverage/waveband` |
| **Coverage.Temporal** | VODataService | `coverage/stc:STCResourceProfile/stc:AstroCoordArea`[1] |
| **Coverage.Spatial** | VODataService | `coverage/stc:STCResourceProfile/stc:AstroCoordArea`[1] |
| **Type.ImageService** | SIA | `capability/imageServiceType` |
| **MaxQueryRegionSize** | SIA | `capability/maxQueryRegionSize` |
| **MaxImageExtent** | SIA | `capability/maxImageExtent` |
| **MaxImageSize** | SIA | `capability/maxImageSize` |
| **MaxFileSize** | SIA | `cabapility/maxFileSize` |
| **MaxRecords** | SIA | `capability/maxRecords` |
| **Verbosity** | SIA | `capability/verbosity` |
| **ImageQueryBaseURL** | VOResource | `capability/interface[@role='std']/accessURL` |

[1] In STC, coverage on the sky and in time are described in an integrated way within the stc:AstroCoordArea element. The notion of the equinox of the observational positions is captured within stc:AstroCoordSystem element.

The SIA Resource extension schema is formally defined in a separate document.